\documentstyle[preprint,eqsecnum,aps,psfig]{revtex}
\begin{document}
\bibliographystyle{unsrt} 
\title{Equation of state of compressed matter: A simple statistical model.} 
\author{L.Delle Site\footnote{Max-Planck-Institute for Polymer
    Science, Theoretical Group, 
PO Box 3148 D-55021 Mainz, Germany;
e-mail address:dellsite@mpip-mainz.mpg.de}}

\address{ Atomistic Simulation Group and Irish Centre for Colloid
Science,\\ School of Mathematics and
Physics,\\ The Queen's University, Belfast BT7 1NN, U.K.}
\date{\today}
\maketitle

\begin{abstract}
We propose a simple approach for studying systems of compressed matter 
based on the Thomas-Fermi statistical model of single atom. The
central point of our work is the development of the concept of
``statistical ionization'' by compression; in simple terms, we
calculate the fraction of electrons within the atom whose positive
energy, due to the compression, exceeds the
negative binding energy electron-nucleus. Next we extend this concept
from a single atom to macroscopic systems and write the corresponding
equation of state. Positive aspects as well as limitations of the
model are illustrated and discussed through all the paper.\\ 
{\bf Author Keywords:} Thomas-Fermi, compressed atoms, statistical
ionization, equation of state.\\
{\bf PACS classification codes:} 05.30Fk, 03.65Sq, 71.10Ca
\end{abstract}
\subsection{Introduction}
The Thomas-Fermi model (see the original work of Fermi \cite{fermi},
or any textbook of condensed matter, e.g. \cite{joachin}) represents a 
simple and powerful tool for the basic investigation of atomic
properties. Although the theoretical framework is highly simplified,
quantitative as well as qualitative results are surprisingly good; 
for example, in the description of compressed atoms, theoretical \cite{fmt} as
well as experimental \cite{ussr} work testify the validity of such 
an approach. In the following part of this work we focus the attention 
on this particular case (atoms under pressure) and develop the
concept of ``statistical ionization'' by compression;
we define the total energy, calculated in a classical way,  
of the compressed atom as a function of the 
distance from the point-like nucleus and using the corresponding 
solution of the Thomas-Fermi equation we are able to define 
the region within the atom characterized by positive
energy, as a consequence we can estimate the
average number of electrons (fraction of electrons) whose 
antibinding energy (again in a classical sense) is 
dominant with respect to the 
binding energy electron-nucleus; the term ``average'' must be interpreted
in the sense that we do not refer to the electrons as 
single particles characterized by specific quantum numbers, but we
refer to them as the results of the process of integration of an
average distribution of charge (i.e. the electron density) over a
certain region of the real space.
Of course, it must be clear, that we do not intend to refer to a such 
ionization mechanism as a realistic one since the electron ionization
does not depend on the distance from the nucleus and 
involves more complicated quantum effects which cannot be considered
by this model; what we intend to do  is simply to estimate 
in a classical 
statistical way the effects of the compression on the
kinetic energy, electron-electron and electron-nucleus interaction. 
Finally from the
single atom, we extend the model to condensed matter systems
introducing, with a statistical approach, the atomistic model into a
multiatomic system. The paper is organized as follows: after a basic
review of the Thomas-Fermi model necessary for the next part of the
paper, we describe the process of
``statistical ionization'' by compression, finally the equation of
state of compressed matter is derived accordingly to the model
illustrated in the previous section. 
\section{The Thomas-Fermi Model}
In this section we will follow the procedure to obtain the
Thomas-Fermi equation reported in reference \cite{joachin}.
The Fermi statistics at zero Kelvin is applied to study the bound
state of an atom characterized by a statistical number of electrons. 
The N electrons of the system are treated as a degenerate
Fermi gas confined in a region of the real space by a spherical
potential $V(r)$ which goes to zero in the limit of $r\to\infty$; it
is supposed that the potential is slowly varying for distances which
are large compared to the De Broglie wavelength, so that in the volume 
where $V(r)$ is approximatively constant we can think that there are
enough electrons to justify the statistical approach of the electrons
as a Fermi gas; moreover the large number of particles allow one to
think that most of them have got an high principal quantum number in
order to justify the application of semiclassical methods. The aim of
the Thomas-Fermi model is to calculate $V(r)$ and the electron density
$\rho(r)$; to do so, these two quantities are related to each other in
the following way. The total energy of an electron can be written as:
\begin{equation}
p^{2}/2m+V(r) 
\end{equation}
with $m$ the mass of an electron , 
since the maximum value of the kinetic energy an
electron can reach is the Fermi energy, the most energetic electrons
of the system will be characterized by a total energy:
\begin{equation}
E_{max}=E_{Fermi}+V(r). 
\end{equation}
It is clear that $ E_{max}$ does not
depend on $r$ because if not all the electrons would migrate in the
region in correspondence of which such a quantity has a minimum.
The Fermi momentum depends on $r$ since
$p^{2}_{Fermi}/2m=E_{max}-V(r)$, and that it can also be expressed as:
\begin{equation}
p_{Fermi}=\hbar(3\pi^{2})^{1/3}(\rho(r))^{1/3} 
\end{equation}
where $\rho(r)=N/V$, 
$N$ total number of electron, $V$ the volume where they are confined 
(see for example \cite{ll}); the combination of these two relations
leads to :
\begin{equation}
\rho(r)=\frac{(2m/\hbar^{2})^{3/2}}{3\pi^{2}}[E_{max}-V(r)]^{3/2}
\end{equation}
in the region classically forbidden where $E_{max}-V(r)<0$ we impose
$\rho(r)=0$. Next the electrostatic potential is defined as:
\begin{equation}
\psi(r)=-\frac{V(r)}{e} 
\end{equation}
with $\psi_{0}=-E_{max}/e$ positive constant 
being $e$ the electron charge. Defining:
\begin{equation}
 \phi(r)=\psi(r)-\psi_{0} 
\end{equation}
it follows that $\rho(r)$ and $\phi(r)$ are related as:
\begin{equation}
\rho(r)=\frac{(2m/\hbar^{2})^{3/2}}{3\pi^{2}}[e\phi(r)]^{3/2}
\end{equation}
for $\phi(r)\ge 0$ and $\rho(r)=0$ for $\phi(r)\le 0$.
Implementing the relations above into the Poisson equation we obtain:
\begin{equation}
\label{poisson}
\frac{1}{r}\frac{d^{2}}{dr^{2}}[r\phi(r)]=\frac {e(2m/\hbar^{2})^{3/2}}{3\pi^{2}\epsilon_{0}}[e\phi(r)]^{3/2}
\end{equation}
for $\phi(r)\ge 0$ \\
and 
\begin{equation}
\frac{1}{r}\frac{d^{2}}{dr^{2}}[r\phi(r)]=0
\end{equation}
for $\phi(r)\le 0$ being $\epsilon_{0}$ the dielectric constant. For $r\to 0$ the predominant term of the
electrostatic potential is due to the nucleus, it follows that:
\begin{equation}
\lim_{r\to 0}r\phi(r)=\frac{Ze}{4\pi\epsilon_{0}} 
\end{equation}
where $Ze$ is the
charge of the nucleus; in addition to the above condition we have to
add the normalization condition $4\pi\int_{0}^{r0}\rho(r)r^{2}dr=N$,
where $r_{0}$ is the radius of the atom. At this point we introduce the 
dimensionless variable and function:
\begin{equation}
r=bx
\end{equation}
\begin{equation}
r\phi(r)=\frac{Ze}{4\pi\epsilon_{0}}\chi(x)
\end{equation}
where $b=\frac{3\pi^{2}}{2^{7/3}}a_{0}Z^{-1/3}$ with $a_{0}$ is the
Bohr radius. The definitions reported above imply the following
relation:
\begin{equation}
\rho(r)=\frac{Z}{4\pi b^{3}}(\chi(x)/x)^{3/2}
\end{equation}
for $\chi(x)\ge 0$ and $\rho(r)=0$ for $\chi(x)< 0$. Finally the
dimensionless equation \ref{poisson} can be written in the form known
as the Thomas-Fermi equation:
\begin{equation}
\label{tf}
\frac{d^{2}}{dx^{2}}\chi(x)=x^{-1/2}\chi(x)^{3/2}
\end{equation}
for $\chi(x)\ge 0$ \\
and 
\begin{equation}
\frac{d^{2}}{dx^{2}}\chi(x)=0 
\end{equation}
for $\chi(x)<0$;\\
the condition in $x=0$ becomes $\chi(0)=1$. It is important
  to notice that equation \ref{tf} is independent from the atomic
  species, in other words is universal and the nature of the atom can
  be reintroduced in the scaling factor; this fact will play an
  important role in the development of our model.
\subsection{Solutions of the Thomas-Fermi equation}
Since equation \ref{tf} is a second order differential 
equation and the boundary condition
in $x=0$ specifies only one of the two required conditions to have a
unique solution, it will exist a class of solutions $\chi(x)$ which
satisfies the condition in $x=0$ and for a specific solution will
depend on the value $\chi^{'}(x)_{x=0}$ (the symbol $^{'}$ means
derivative with respect to $x$). The general solution of the
Thomas-Fermi equation can be classified in three categories:\\
(1) Asymptotic solution obtained for $\chi^{'}(x)_{x=0}\approx
-1.576...$\\
(2) Solution which goes to zero for finite values of $x$, i.e
$x=x_{0}$, for $\chi^{'}(x)_{x=0}<-1.576...$\\
(3) Solution which diverges for large $x$, for
$\chi^{'}(x)_{x=0}>-1.576...$.\\
We do not report a pictorial description of these solutions, since it
can be found in details in reference \cite{joachin}.
The physical meaning of the three categories of solutions can be
understood examining the normalization condition.
In fact:
\begin{equation}
N=Z\int_{0}^{x_0}x^{1/2}\chi(x)^{3/2}dx 
\end{equation}
leads to
\begin{equation}
N=Z[x\chi^{'}(x)-\chi(x)]_{0}^{x_0} 
\end{equation}
and finally to :
\begin{equation}
\frac{N-Z}{Z}=x_{0}\chi^{'}(x_{0})-\chi(x_{0}).
\end{equation}
In the asymptotic case we have that $\chi^{'}(x)$ goes to zero 
as $\chi(x)$ for large $x$ so we have the neutral atom with the
infinite radius; in the second case we have that $\chi(x_{0})=0$ but $
\chi^{'}(x_{0})x_{0}\neq 0$ so we obtain $N-Z<0$ being
$\chi^{'}(x_{0})<0$, i.e positive ions; in the third case
which is the one particularly relevant for our work, we have:
\begin{equation}
\chi^{'}(x_{0})x_{0}-\chi(x_{0})=0
\end{equation}
where $x_{0}$ is smaller than the radius of the neutral atom (infinite 
as stated before) so that such a solution is interpreted as the
solution describing compressed neutral atoms.
\section{Statistical Ionization by compression}  
In this section we will illustrate the concept of statistical ionization
by compression in physical as well as mathematical terms.
First, we make the ideal picture of the atom as composed of 
concentric shells of infinitesimal thickness around the point-like
nucleus, in this approach the total energy
within the atom at a certain distance from the nucleus is represented 
by the sum of the energy of each single infinitesimal shell contained in
the volume
corresponding to that distance, or equivalently, for a continuum of
shells, 
by the integral over that volume. 
Next, we write the total energy of the atom 
in terms of solutions of the Thomas-Fermi 
equation for compressed atoms within the idealized shells' approach; 
this simply leads to a semianalytical function of the distance from the
nucleus.
At this point it we can be noticed that the distance from the nucleus 
at which this function has got a minimum
corresponds to the shell characterized by a null contribution to the total
energy;
from a physical point of view this means that this shell is characterized
by the exact balance between binding and antibinding energy or in other
words is the distance at which the attractive potential between the nucleus
and the electrons starts to
be less dominant than the electron-electron and kinetic contributions.
Once this distance has been found, it is automatically defined a
region 
within the atom, characterized by a positive energy; 
integrating over this region, one
obtains the average number of electrons whose binding interaction with the
nucleus can be considered negligible, i.e. they can be, with a good
approximation, considered free.    
As stated before, this process is not meant to reproduce a realistic
ionization, but 
it represents a simple mechanism which in terms of classical 
interactions between the
physical elements of the system helps one to picture out the balancing
process between the reduced volume available to the atom and the
topological readjustment of the electron density; the average number of
electrons considered free must be interpreted as a fraction
of electrons which can be represented as
a non interacting  electron gas. It must be underlined that this is not
obvious, since the Thomas-Fermi model is based on the hypothesis of non
interacting electron gas  and
we use this description to define an electron-electron interaction and
calculate the fraction of electrons free from the nucleus; the argument 
which one can use to solve this critical point is that once the electron
distribution is defined via the Thomas-Fermi equation as a further step the
electrostatic interaction between electrons can be calculated as a usual
classical self-interacting charged sphere and that the initial hypothesis
of non interacting electrons is just
a simplification to obtain at a first step a reasonable atomistic electron
distribution;
at the same time the ``ionized'' electrons can be interpreted as 
non interacting
fermionic electron gas in a less approximate way than the total number of
electrons considered initially in the Thomas-Fermi model since these former
do not feel the nucleus attraction. It must be clear , as stated also before, 
that a realistic
process of ionization cannot depend 
on the distance of the electrons from the nucleus, in our model this simply
represents a sort of classical way to describe the process and is
related to the semiclassical
and statistical nature of the Thomas-Fermi model; this means 
that it makes sense
within a semiclassical framework but not for example 
in a proper quantum treatment.
\subsection{The Total Energy Function or Ionization Function}
In this section we write the total energy of the atom as a distance from
the nucleus
using the solutions for the compressed atoms. In a classical approach, the
total energy of the electrons ``located'' at distance $R$ from the nucleus
is (see also reference \cite{shap-teul}):
\begin{equation}
E(R)_{total}=E(R)_{kinetic}+E(R)_{electron-electron}+E(R)_{electron-nucleus}
\end{equation}
\subsubsection{The Kinetic Energy}
The kinetic energy is the energy of a fermionic gas in a sphere 
at zero Kelvin:
\begin{equation}
E(R)_{kinetic}=\int_{0}^{R}4\pi r^{2}\int_{0}^{p_{Fermi(r)}}\frac{4\pi
  p^{4}}{m\hbar^{3}}dp
\end{equation}
this can be written in terms of Thomas-Fermi adimensional
quantities as:
\begin{equation}
E_{kinetic}(s)=\frac{3Z^{2}e^{2}}{5
  b}\int_{0}^{s}\chi(x)^{5/2}x^{-1/2}dx
\end{equation}
where $s=R/b$ and $x=r/b$ being $b$ the same as defined before.
This integral can be simplified in a useful 
semianalytical form (what we mean 
is that the final form is analytical in $\chi$ and $\chi^{'}$, but is
globally semianalytical since $\chi$ and  $\chi^{'}$ are numerical
solutions, as functions of the dimensionless distance from the
nucleus).
Considering that $\chi^{''}(x)=\chi^{3/2}(x)/x^{1/2}$, where
$\chi^{''}$ is the second derivative with respect to $x$, the integral 
can be rewritten as:
\begin{equation}
\label{integ1}
\int_{0}^{s}\chi(x)^{5/2}x^{-1/2}dx= \int_{0}^{s}\chi^{''}d\chi
\end{equation}
which is equivalent to:
\begin{equation}
\int_{0}^{s}\chi^{''}d\chi=\int_{0}^{s}\chi d\chi^{'}
\end{equation}
taking into account that $\chi(0)=1$ and
integrating by parts, we obtain:
\begin{equation} 
\int_{0}^{s}\chi
d\chi^{'}=\left[\chi(x)\chi(x)^{'}\right]^{s}_{0}-\int_{0}^{s}\chi^{'} d\chi.
\end{equation}
The the second term of the right side of the 
previous equation can be expressed as:
\begin{equation}
-\int_{0}^{s}\chi^{'} d\chi=-\int_{0}^{s}(\chi^{'})^{2}dx
\end{equation}
which, using again the integration by parts and the properties of the
Thomas-Fermi equation, becomes:
\begin{equation}
-\int_{0}^{s}(\chi^{'})^{2}dx=\left[-(\chi(x)^{'})^{2}x\right]^{s}_{0}+
\left[\frac{4}{5}\chi(x)^{5/2}x^{1/2}\right]^{s}_{0}-\frac{2}{5}
\int_{0}^{s}\chi(x)^{5/2}x^{-1/2}dx.
\end{equation}
It follows that:
\begin{equation}
\int_{0}^{s}\chi(x)^{5/2}x^{-1/2}dx=\left[\chi(x)\chi(x)^{'}\right]^{s}_{0}-
\left[(\chi(x)^{'})^{2}x\right]^{s}_{0}+
\left[\frac{4}{5}\chi(x)^{5/2}x^{1/2}\right]^{s}_{0}-\frac{2}{5}
\int_{0}^{s}\chi(x)^{5/2}x^{-1/2}dx.
\end{equation}
Finally:
\begin{equation}
E(s)_{kinetic}=\frac{3Z^{2}e^{2}}{7b}\left[\chi(s)\chi^{'}(s)-\chi^{'}(0)-(\chi^{'}(s))^{2}s+\frac{4}{5}(\chi(s))^{5/2}s^{1/2}\right].
\end{equation}
\subsubsection{The Electron-Nucleus Interaction}
In this part, we write the electron-nucleus interaction in terms of
the Thomas-Fermi quantities.
The electron-nucleus interaction is written as:
\begin{equation}
E_{e-n}(R)=-Ze^{2}\int_{0}^{R}4\pi r^{2}\frac{\rho(r)}{r}dr
\end{equation}
and within the Thomas-Fermi approach becomes:
\begin{equation}
E_{e-n}(s)=-\frac{Ze^{2}}{b}\int_{0}^{s}[\chi(x)]^{3/2}x^{-1/2}dx=-\frac{Ze^{2}}{b}\int_{0}^{s}\chi(x)^{''}dx.
\end{equation}
Finally, integrating by parts we obtain:
\begin{equation}
E_{e-n}(s)=\frac{Ze^{2}}{b}[\chi^{'}(s)-\chi^{'}(0)].
\end{equation}
\subsubsection{The Electron-Electron Interaction}
In the same fashion of the previous calculations, we repeat the
procedure for the electron-electron interaction.
\begin{equation}
E_{e-e}({\bf R})= \frac{e^{2}}{2}\int_{V}\rho({\bf r})d{\bf
  r}\int_{V_0}\frac{\rho({\bf r}^{'})}{|{\bf r}-{\bf r}^{'}|}d{\bf
  r}^{'}
\end{equation}
the bold letters indicate that the related quantity is in Cartesian
coordinates, $V$ is a varying spherical region within the atom while
$V_{0}$ is the total volume of the atom. 
This integral can be simplified in terms of Thomas-Fermi quantities in 
the following way (see also
reference \cite{shap-teul}):
\begin{equation}
E_{e-e}({\bf
  s})=\frac{1}{2}\frac{Z^{2}e^{2}}{b}\int_{0}^{s}[\chi(x)]^{3/2}
x^{1/2}dx\left[\frac{1}{x}\int_{0}^{x}[\chi(x^{'})]^{3/2}(x^{'})^{1/2}+
\int_{0}^{s_0}[\chi(x^{'})]^{3/2}(x^{'})^{1/2}\right]
\end{equation}
where $s_{0}=R_{0}/b$, being $R_{0}$ the total radius of the atom.
Applying the Thomas-Fermi equation $\chi^{''}=\chi^{3/2}/x^{1/2}$, and 
integrating by parts we obtain:
\begin{equation}
\label{ee}
E_{e-e}({\bf
  s})=\frac{1}{2}\frac{Z^{2}e^{2}}{b}\int_{0}^{s}[\chi(x)]^{3/2}x^{1/2}dx[-\chi(x)/x+\chi(0)/x+\chi^{'}(x_{0})]
\end{equation}
and this gives:
\begin{equation}
\int_{0}^{s}[\chi(x)]^{5/2}x^{-1/2}dx+\chi(0)\int_{0}^{s}[\chi(x)]^{3/2}x^{-1/2}dx+\chi^{'}(x_{0})\int_{0}^{s}[\chi(x)]^{3/2}x^{1/2}dx
\end{equation}
the first term is equivalent to the integral of the kinetic energy,
the second is equivalent to the integral of the electron-nucleus
energy multiplied by $2$, the integral of the third term can be solved
in the same way as the two inner integrals of the electron-electron
energy in equation \ref{ee}; it follows that the final form for the
electron-electron energy in units of $Z^{2}e^{2}/b$ is:
\begin{equation}
\begin{array}{cc}
E_{e-e}(s)=\frac{1}{14}\chi(s)\chi^{'}(s)+\\
\frac{3}{7}\chi^{'}(0)-\frac{1}{14}s(\chi^{'}(s))^{2}+\\
\frac{2}{35}s^{1/2}\chi(s)^{5/2}-\frac{1}{2}\chi^{'}(s)+\frac{1}{2}s
\chi^{'}(s_{0})\chi{'}(s)-\frac{1}{2}\chi^{'}(s_{0})\chi(s)\end{array}.
\end{equation} 
\subsubsection{Final Form of the Ionization Function and Qualitative Numerical
  Results}
In this part we give the final form of the total energy, and show a 
pictorial representation of its curve for three different degrees of
compression. Combining the results of the integration performed in the 
previous sections we can write the total energy (in arbitrary units) as:
\begin{equation}
\begin{array}{cc}E_{tot}(x)=\frac{1}{2}\chi(x)\chi^{'}(x)-
\frac{1}{2}[\chi^{'}(x)]^{2}x+
\chi^{'}(0)+\frac{14}{35}[\chi(x)]^{5/2}x^{1/2}\\
-\frac{3}{2}\chi^{'}(x)+
\frac{1}{2}x\chi^{'}(x_{0})\chi^{'}(x)-\frac{1}{2}\chi^{'}(x_{0})\chi(x)\end{array}
\end{equation}
where we formally replaced $s$ with $x$, previously used as the
integration variable.
The average number of ``ionized'' electron is given by:
\begin{equation}
N_{i}=Z\int_{x_i}^{x_0}x^{1/2}[\chi(x)]^{3/2}dx=Z[\chi(x_{i})-x_{i}\chi^{'}(x_{i})]
\end{equation}
$x_{i}$ is the dimensionless distance at which the minimum of the
ionization function is located.
We studied numerically the behavior of the ionization function 
for three different degrees of compression, which in mathematical terms 
means that we used three different solutions of the Thomas-Fermi
equation corresponding to three different initial conditions on $\chi^{'}(0)$;
figure \ref{picture}, shows the qualitative behavior of such a
function; indeed a minimum is found, and the corresponding
dimensionless distance can be used for calculating the average number
of ``ionized'' electrons. As we expect for very
compressed atoms ($\chi^{'}(0)$ less negative) the distance at which
the positive contributions to the energy start to be dominant 
(with respect to the less compressed atoms) 
is shorter, or in simple terms, the average number of ``ionized''
electrons is larger. This of course simply shows that the model
does not present an unphysical behavior and is qualitatively
reasonable but, as stated before, it should not be directly
compared with specific ionization studies.
\section{Equation of State of Systems under Pressure}
In this part we will apply the model developed before to write the
equation of state of systems under pressure. The original work of the
ionization by compression \cite{luigitesi} was developed for
describing astrophysical systems at high density and consequently at
extreme pressure (white dwarfs). In this case  
the usual basic approach used in
literature (see for example \cite{ll}) 
consists in considering the atoms fully ionized and the
equation of state is written in the approximation of the electrons a
perfect Fermi gas in the ground state, while
the nuclei are not suppose to contribute to the pressure. In
our case the approximation is less crude since we do estimate the
average number of electrons per atom which at a given pressure or
equivalently at a 
given density of matter of the system 
(in terms of the inverse of the volume of the atom) can be considered
free from the nucleus attraction so that for them it is more appropriate to
apply the properties of the perfect Fermi gas. 
In a large system of a single species of atoms we can imagine the
compressed atom as the one described by the Thomas-Fermi model and the 
pressure experienced by the atom is the same the itself atom produces on the
system in a situation of equilibrium. 
It follows that if we consider all the electrons of the atom as a
Fermi gas, the equation of state is:
\begin{equation}
P=3\pi^{4/3}\frac{\hbar^{2}}{5m_{e}}\left[\frac{N}{V}\right]^{5/3}    
\end{equation}
where $N$ is the total number of electrons of the atom and $V$ the
corresponding volume;
this formula is not fully justified when we use our approach of partial
ionization since in this case the equation takes the form:
\begin{equation}
\label{la15}
P=3\pi^{4/3}\frac{\hbar^{2}}{5m_{e}}\left[\frac{N_{i}(V)}{V}\right]^{5/3}.     
\end{equation}
where $N_{i}(V)$ is the average number of ``ionized'' electrons.
The approximation of considering the ``ionized'' electrons as a Fermi gas, 
should be introduced in the equation of state in a statistical way,
since the pressure is a statistical quantity and must be calculated
using a statistically number of electrons; the number of 
``ionized'' electrons for single atom is clearly not fulfilling such
a requirement. To do so, we use the following procedure, we take  a
volume inside the many-atom system containing a large number of
atoms $N_{a}$ which we consider under the same physical conditions. It 
follows that the
equation of state can be written as:
\begin{equation}
P=3\pi^{4/3}\frac{\hbar^{2}}{5m_{e}}
\left[\frac{N_{a}N_{i}(V)}{N_{a}V}\right]^{5/3}
\end{equation}
in this case the total number of free electrons contributing to the
pressure is $N_{a}N_{i}(V)$ and the total volume in which they are
confined is $N_{a}V$;
the above equation obviously reduces to equation \ref{la15}.
In simple terms, the ``ionized'' electrons are considered 
a Fermi gas in the ground state while the
remaining  electrons are approximate as a frozen core containing also 
the nucleus , in the same fashion, with due differences, of the
pseudopotential approach used in modern first principles calculations
of condensed systems. The equation obtained can be extended to any
system, regardless of the size; of course we expect that large
systems at relatively high pressure represent better the statistical
framework required by the approximations done, this is the reason we developed
the procedure keeping in mind the case of astrophysical objects. 
\section{Discussion and Conclusions}
As stated all over the paper, this is a very simple model,
we do not expect to obtain valid precise quantitative results; however 
when used in a semiclassical statistical framework, it could give
valid indications. Certainly the approach described in reference
\cite{ll} for large systems under pressure represents a crude 
approximation; the same can be said for our model, but once the
approximation of ``frozen core''
 electrons is accepted, the ``ionized'' electrons 
well represent an ideal Fermi gas. Of course when the pressure
becomes extremely high, relativistic effects as well as nuclear
processes (e.g. inverse $\beta$-decay) become relevant and the approximation 
of non-relativistic Fermi gas is not justified anymore; this means
that our model is valid up to a certain degree of
compression. Moreover, the model can be applied in principle to any
species of atoms by simply scaling all the quantities by the factor
$Z$,
this is a relevant property which makes the ionization function
universal. 
However when more sophisticated models
of the Thomas-Fermi equation are used the universality of the
description is lost (for
example when electronic exchange and correlation effects are included
\cite{fmt}).  
We expect the model to be less valid  for ordinary 
condensed systems under pressure where do not exist fully idealized
statistical conditions and the details of a quantum description become 
relevant. Nevertheless also in this case our approach can be
useful to have a first estimate of the compression process; in
particular in many self consistent first principles calculations where the
pseudopotential approach is used, our model could be used to estimate
the number of ``ionized'' electrons so that at the first step of the
self-consistency they can be represented by a plane-wave electron
wavefunction  while the
remaining electrons can be approximate by a frozen core or by closed
shell orbitals centered on the nucleus; this could speed up the
convergence of the self-consistent process. In conclusion, we think
that due to the its simplicity and feasibility, our model, when is
used in a proper context, represents a
useful tool for the basic investigation of statistical properties of compressed systems.
\section{Later correction:Physica A295 562 (2001)}
Eq. (3.4) right hand side should be\\
$\int_{0}^{s}\chi^{''}\chi dx$ instead of $\int_{0}^{s}\chi^{''}d\chi$.\\

The third integral on the right hand side of Eq.(3.15) should be\\
$\int_{x}^{s_{0}}[\chi(x')]^{3/2}(x')^{-1/2}$ instead of $\int_{x}^{s_{0}}[\chi(
x')]^{3/2}(x')^{1/2}$.\\
Eq. (3.17), the sign of the first integral is $-$.\\
Eq. (3.18), for the first and third term on the right hand side is $-5/14$ inste
ad of $1/14$ the
multiplicative factor, for the fourth term is $-10/35$ instead of $2/35$, and th
e fifth
term is minus instead of $+$.\\
Eq. (3.19), for the first and second term on the right hand side is $1
/14$ instead of $1/2$ the
multiplicative factor, for the fourth term is $2/35$ instead of $14/35$, and for
 the fifth
term is $-1/2$ instead of $-3/2$.
The rest remains unaltered. 

\subsection{Acknowledgments}
 I thank EPSRC
for financial support (grants GR/K20651 and GR/L08427) and the IFI for
financial support for the Irish Centre for Colloids and Biomaterials.

\begin{figure} \caption{The total energy of the atom (or equivalently
    the ionization function as we also address to it) as a function
of the dimensionless distance from the nucleus for three different
degrees of compression; the dimensions
 are arbitrary since only the qualitative behavior is relevant;
both qualitative as well as quantitative behavior do not have a
physical sense for a distance which is larger than the atom's radius,
accordingly to the interpretation of the solutions of the Thomas-Fermi 
equation for compressed atoms.
\label{picture}} \end{figure}

\begin{figure}
\centerline{\psfig{figure=good-tf.ps,width=1.0\textwidth,angle=-90}}
\end{figure}

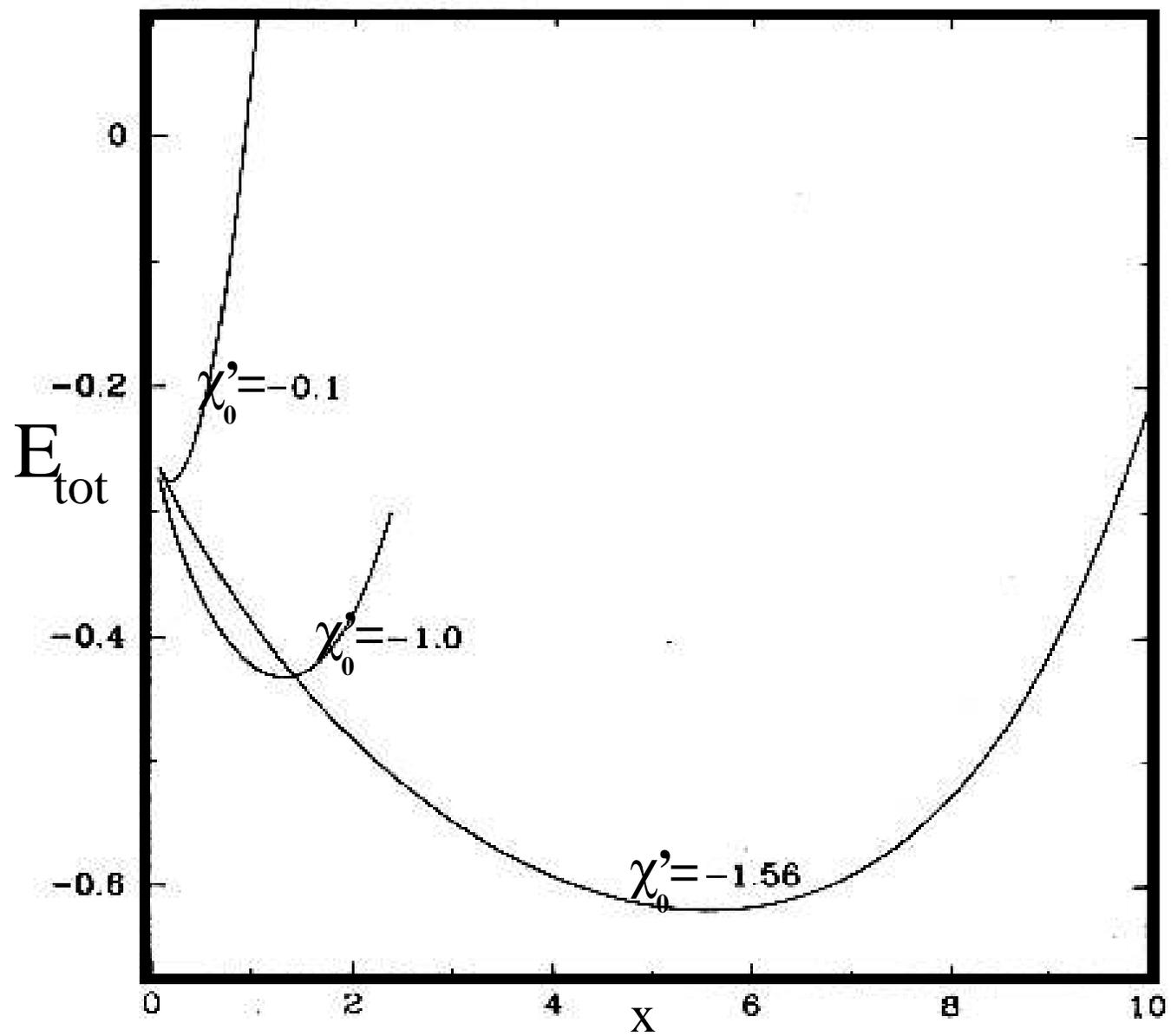
\end{document}